# Why High-Spin Prograde MAD States Are Not Realized for AGN in Nature: Consequences for AGN Phenomenology

David Garofalo

Department of Physics, Kennesaw State University, USA

Abstract

Quasar-like and other bright active galactic nuclei are typically radio quiet despite strong evidence that many host rapidly spinning black holes. In contrast, general relativistic magnetohydrodynamic (GRMHD) simulations show that high-spin prograde black holes in magnetically arrested disk (MAD) states produce the most powerful relativistic jets. We show that these results are not in conflict but instead correspond to physically distinct and mutually limiting evolutionary pathways. If accretion proceeds through radiatively efficient thin disks, then as the innermost stable circular orbit decreases with increasing prograde spin, radiative output and radiation pressure rise sharply, driving strong disk winds that remove the conditions required for effective jet formation. In this branch, high-spin prograde systems are therefore expected to be jet-suppressed. Alternatively, if accretion transitions to radiatively inefficient flows capable of producing MAD states, then the spin-up to near-maximal prograde rotation proceeds on timescales that exceed the age of the universe for the relevant systems, preventing the realization of the extreme high-spin prograde MAD configurations explored in GRMHD simulations. Within this framework, powerful jets need only be produced efficiently in counterrotating systems, not maximized there, preserving the time-dependent spin evolution central to the counterrotation versus corotation paradigm of Garofalo et al. (2010). We thus reconcile GRMHD results with observations and provide a unified interpretation of the radio-loud/radio-quiet dichotomy and the cosmic evolution of powerful radio sources.

## Introduction

Relativistic jets from accreting black holes are among the most energetic phenomena in the universe and are widely understood to arise from the extraction of rotational energy mediated by large-scale magnetic fields. The Blandford–Znajek mechanism (Blandford & Znajek 1977) provides the theoretical foundation for this process, predicting that jet power scales with both black hole spin and the magnetic flux threading the event horizon. Over the past two decades, general relativistic magnetohydrodynamic simulations have confirmed and extended this picture, demonstrating that powerful jets can be launched from accreting black holes and that jet efficiency increases with spin, particularly in magnetically arrested disk (MAD) states (e.g., McKinney 2006; Tchekhovskoy, Narayan & McKinney 2011).

Within this framework, the most powerful jets are associated with rapidly spinning black holes accreting in prograde configurations with large magnetic flux accumulation. These results are often interpreted as implying that high-spin black holes should generically be associated with strong radio emission. However, observational studies of active galactic nuclei (AGN) present a more complex picture. Quasar-like and other bright AGN are predominantly radio quiet (e.g., Kellermann et al. 1989; White et al. 2000), despite mounting evidence that many such systems host rapidly spinning black holes (e.g., Reynolds 2014). At the same time, powerful radio-loud AGN, including FRII quasars, exhibit a distinct cosmic evolution, dominating the high-redshift universe (e.g., Dunlop & Peacock 1990; Best et al. 2014).

These observational trends have motivated alternative frameworks in which jet production depends not only on spin magnitude but also on the orientation of the accretion flow relative to the black hole spin. In particular, the counterrotation–corotation paradigm developed in Garofalo, Evans & Sambruna (2010), building on earlier work (Garofalo 2009; Reynolds, Begelman & Garofalo 2006), provides a unified interpretation of AGN phenomenology. In its original formulation, this framework emphasized the possibility that retrograde (counterrotating) accretion could produce more powerful jets than prograde accretion due to enhanced magnetic flux accumulation. However, it has since become clear that this requirement is unnecessary. The paradigm does not depend on retrograde systems producing the most powerful jets, but rather on their ability to produce powerful jets and to enable the time-dependent spin evolution that links different classes of AGN.

In this picture, retrograde accretion is associated with the production of powerful jets, while prograde accretion, particularly in thin disks, is linked to systems in which strong radiative output is accompanied by suppression of jet formation. This framework naturally accounts for the radio-loud/radio-quiet dichotomy, the observed anti-correlation between jets and disk winds (e.g., Tombesi et al. 2010; Reeves et al. 2009), and the cosmic evolution of radio sources.

Despite its explanatory power, the counterrotation–corotation paradigm is often perceived to be in tension with results from numerical simulations, which emphasize the efficiency of jet production in high-spin prograde MAD states. This apparent discrepancy is typically framed as a conflict between the role of prograde accretion in simulations and retrograde accretion in phenomenological models. However, this interpretation fails to account for several fundamental physical constraints that limit which accretion states are actually realized in nature.

First, the accretion flows associated with bright AGN are expected to be geometrically thin. As the innermost stable circular orbit decreases for increasing prograde spin, radiative efficiency and radiation pressure increase sharply, leading to strong disk outflows that modify the accretion environment and remove the conditions required for effective jet formation. In this regime, high-spin prograde systems are not expected to produce powerful jets.

Second, magnetically arrested disk states are not initial conditions but the result of an evolutionary process in which large-scale magnetic flux accumulates at the black hole over time. The formation of a MAD state therefore requires sustained inward transport of magnetic flux under conditions that do not strongly expel it. In radiatively efficient thin disks, strong radiation-driven outflows act against this process, while in radiatively inefficient flows capable of supporting MAD states, the timescales required to spin black holes up to near-maximal prograde rotation are comparable to or exceed the age of the universe for the relevant systems. As a result, the high-spin prograde MAD configurations that maximize jet power in simulations are not physically realized.

Within this framework, there is no requirement that counterrotation produce the most powerful jets. It is sufficient that powerful jets are efficiently generated in counterrotating systems, allowing the time-dependent spin evolution central to the counterrotation–corotation paradigm to proceed. We therefore show that numerical simulations and the gap paradigm are not in conflict but instead describe different regions and evolutionary stages of black hole accretion. This provides a unified interpretation of jet production and the radio-loud/radio-quiet dichotomy in active galactic nuclei.

## 2. Jet Production in Numerical Simulations of Accreting Black Holes

General relativistic magnetohydrodynamic simulations have established a robust framework for understanding jet production from accreting black holes. Within this framework, relativistic jets are powered by the extraction of rotational energy through large-scale magnetic fields, in agreement with the Blandford–Znajek mechanism. A key result emerging from these simulations is that jet power depends strongly on both black hole spin and the magnetic flux threading the horizon. In particular, jet power can be expressed in the form

$$P_{\mathrm{jet}} \propto \Phi^2 \Omega_H^2,$$

where $\Phi$ is the magnetic flux accumulated on the black hole and $\Omega_H$ is the angular velocity of the horizon. This scaling has been verified across a wide range of numerical studies and provides a quantitative link between accretion flow properties and jet energetics.

A central result of these simulations is the identification of magnetically arrested disk (MAD) states, in which the accumulation of large-scale poloidal magnetic flux becomes sufficiently strong to influence the inner accretion flow. In this regime, magnetic pressure approaches the ram pressure of the inflowing gas, leading to a saturation of magnetic flux on the horizon and highly efficient jet production. Simulations consistently show that the most powerful jets are produced in MAD states around rapidly spinning black holes.

Within this context, prograde accretion configurations are found to be particularly efficient at producing jets when combined with high spin and strong magnetic flux, and the highest jet efficiencies in simulations are typically obtained for rapidly spinning black holes in prograde MAD states. However, it is important to emphasize that powerful jets are not restricted to prograde configurations. Numerical studies have demonstrated that retrograde accretion can also produce substantial jet power, particularly in magnetically arrested or near-arrested states (e.g., McKinney et al. 2012 and related work). Thus, while prograde configurations can maximize jet efficiency under certain conditions, the ability to produce powerful jets is not exclusive to them.

The accretion flows modeled in these simulations are typically geometrically thick and radiatively inefficient, corresponding to conditions often associated with advection-dominated accretion flows (ADAFs). In the context of the counterrotation–corotation paradigm, such flows are not generic but instead represent a later stage in the evolution of accreting black hole systems in AGN. As a result, the MAD states realized in simulations should be understood as describing a specific regime of accretion rather than a universal outcome.

Taken together, these results define a regime in which rapidly spinning black holes accreting in magnetically arrested, radiatively inefficient flows generate powerful relativistic outflows. In what follows, we examine how this regime relates to the broader evolutionary picture of accreting black holes and whether the extreme high-spin prograde MAD configurations that maximize jet power in simulations are realized in nature.

## 3. Observational Constraints from Active Galactic Nuclei

Observations of active galactic nuclei (AGN) reveal a pronounced dichotomy in radio properties. While a minority of systems exhibit powerful, extended radio emission associated with relativistic jets, the majority of quasar-like and other bright AGN are radio quiet (e.g., Kellermann et al. 1989; White et al. 2000). This dichotomy persists across a wide range of luminosities and redshifts and is not trivially explained by differences in black hole mass or accretion rate alone.

The radio-loud population is characterized by large-scale radio structures, including the classical double-lobed morphology of Fanaroff–Riley type II (FRII) sources, while radio-quiet systems lack such prominent jet signatures. The existence of this dichotomy suggests that jet production is not a universal outcome of black hole accretion but instead depends sensitively on additional physical parameters.

Powerful radio-loud AGN exhibit a strong cosmic evolution. FRII quasars are observed to dominate the radio-loud population at high redshift, with their space density peaking at $z \sim 2$ and declining toward the present epoch (e.g., Dunlop & Peacock 1990; Best et al. 2014). In contrast, lower-power radio sources and radio-quiet AGN become increasingly prevalent at lower redshift.

This redshift dependence implies that the most powerful jets are associated with earlier stages of black hole and galaxy evolution. Any viable model of jet production must therefore account not only for the existence of powerful jets, but also for their systematic evolution with cosmic time. In particular, the decline of powerful FRII sources toward low redshift suggests an underlying evolutionary sequence in which the conditions for producing strong jets change over time.

Further insight into the physics of jet production is provided by the observed relationship between jets and disk-driven outflows. X-ray observations of AGN have revealed the presence of powerful winds, including ultra-fast outflows with velocities approaching a significant fraction of the speed of light (e.g., Reeves et al. 2009; Tombesi et al. 2010). These winds are commonly associated with systems lacking strong radio jets.

More generally, an anti-correlation is observed between the presence of strong winds and the production of powerful relativistic jets. Systems exhibiting prominent outflows tend to be radio quiet, while those with strong jets often show weaker signatures of disk winds. This behavior is pronounced in X-ray binaries, where transitions between accretion states are accompanied by a switch between jet-dominated and wind-dominated regimes.

The existence of this anti-correlation indicates that jet production and disk outflows are competing channels for the release of accretion energy. Any theoretical framework must therefore account for both the conditions under which jets are efficiently produced and those under which they are suppressed.

Taken together, these observational constraints place strong requirements on models of jet production in AGN. First, any viable framework must explain why the majority of bright, quasar-like systems are radio quiet despite evidence for rapidly spinning black holes. Second, it must account for the strong cosmic evolution of powerful radio sources, particularly the prominence of FRII quasars at high redshift. Third, it must incorporate the observed anti-correlation between jets and disk winds.

These considerations suggest that jet production is not determined solely by black hole spin or magnetic flux but instead depends on the broader accretion environment and its evolution over time. In particular, the data require a framework in which both the production of powerful jets and their suppression arise naturally under different physical conditions.

In the following section, we revisit the counterrotation–corotation paradigm in light of these observational constraints and clarify the role of spin orientation and accretion state in shaping the phenomenology of active galactic nuclei.

## 4. The Counterrotation–Corotation Paradigm Revisited

The counterrotation–corotation paradigm developed in Garofalo, Evans & Sambruna (2010), building on earlier work (Garofalo 2009; Reynolds, Begelman & Garofalo 2006), provides a framework in which the orientation of the accretion flow relative to the black hole spin plays a central role in determining the properties of active galactic nuclei. In this picture, retrograde (counterrotating) accretion is associated with the production of powerful relativistic jets, while prograde accretion, particularly in thin disks, is linked to systems in which jet formation is suppressed.

In its original formulation, the paradigm was motivated in part by the possibility that retrograde accretion could lead to enhanced magnetic flux accumulation on the black hole, thereby producing more powerful jets than prograde configurations. This idea provided a mechanism for associating the most powerful radio-loud systems with retrograde accretion, while naturally connecting prograde systems with radio-quiet AGN.

Subsequent developments indicate that this requirement is not necessary. The ability of retrograde configurations to produce powerful jets does not depend on their exceeding the jet power achievable in prograde configurations. Rather, it is sufficient that retrograde accretion efficiently produces powerful jets under realistic astrophysical conditions. In addition, the association of counterrotation with more powerful jets was motivated in part by observations that FRI jets are weaker and are naturally linked to prograde accretion in the paradigm. However, this distinction is now understood to arise from a combination of environmental factors and the properties of the central engine, both of which are shaped by the evolutionary transition from counterrotation to corotation.

In this revised view, the essential role of retrograde accretion is not to maximize jet power, but to provide a regime in which powerful jets are naturally produced and sustained. This distinction is critical. The paradigm does not require that retrograde systems define the upper bound of jet efficiency, but only that they occupy a region of parameter space in which strong jet production is possible.

This shift removes the need to invoke systematic differences in magnetic flux between retrograde and prograde configurations as the primary driver of the radio-loud/radio-quiet dichotomy. Instead, it places emphasis on the broader physical conditions associated with different accretion regimes and spin orientations.

A key feature of the counterrotation–corotation paradigm is its inherently evolutionary character. In this framework, black hole systems evolve through a sequence of accretion states driven by changes in both spin and accretion orientation. Retrograde accretion leads to a spin-down of the black hole, driving the system toward zero spin and, under continued accretion, into a prograde configuration. This transition naturally connects phases of powerful jet production with later stages in which jet activity is reduced or suppressed. Such an evolutionary sequence provides a natural explanation for the observed decline in the space density of powerful FRII quasars from high redshift to the present epoch.

The revised paradigm provides a unified interpretation of the observational constraints outlined in Section 3. The radio-loud/radio-quiet dichotomy emerges naturally from the distinction between accretion configurations and their associated environments. Retrograde systems are associated with powerful jet production, while high spin prograde thin-disk systems correspond to the radio-quiet population.

At the same time, the observed anti-correlation between jets and disk winds follows directly from the different physical conditions associated with these configurations. Systems with strong radiative output and disk-driven outflows are linked to prograde thin disks, while those capable of sustaining powerful jets correspond to conditions in which such outflows are weak or absent.

Finally, the cosmic evolution of radio-loud AGN is naturally explained by the evolutionary sequence inherent in the paradigm, in which retrograde systems dominate at early times and transition to prograde configurations at later stages.

Crucially, this perspective allows for full compatibility with simulation results demonstrating that prograde configurations can, in principle, produce highly efficient jets under magnetically arrested conditions. However, when the evolutionary nature of accretion and the physical constraints associated with different accretion states are considered, a more restrictive picture emerges. The configurations that maximize jet power in simulations—namely, rapidly spinning black holes in prograde MAD states—require conditions that are not simultaneously satisfied in realistic astrophysical systems.

The resulting framework is therefore not only internally consistent, but also capable of accommodating a wide array of observations, including the radio-loud/radio-quiet dichotomy, the jet–wind anti-correlation, and the cosmic evolution of powerful radio sources. Taken together, these observations imply that high-spin prograde MAD configurations are not realized in nature. Rather, they represent limiting cases that define the upper bounds of jet production in simulations but are excluded by the combined constraints of accretion physics and black hole evolution.

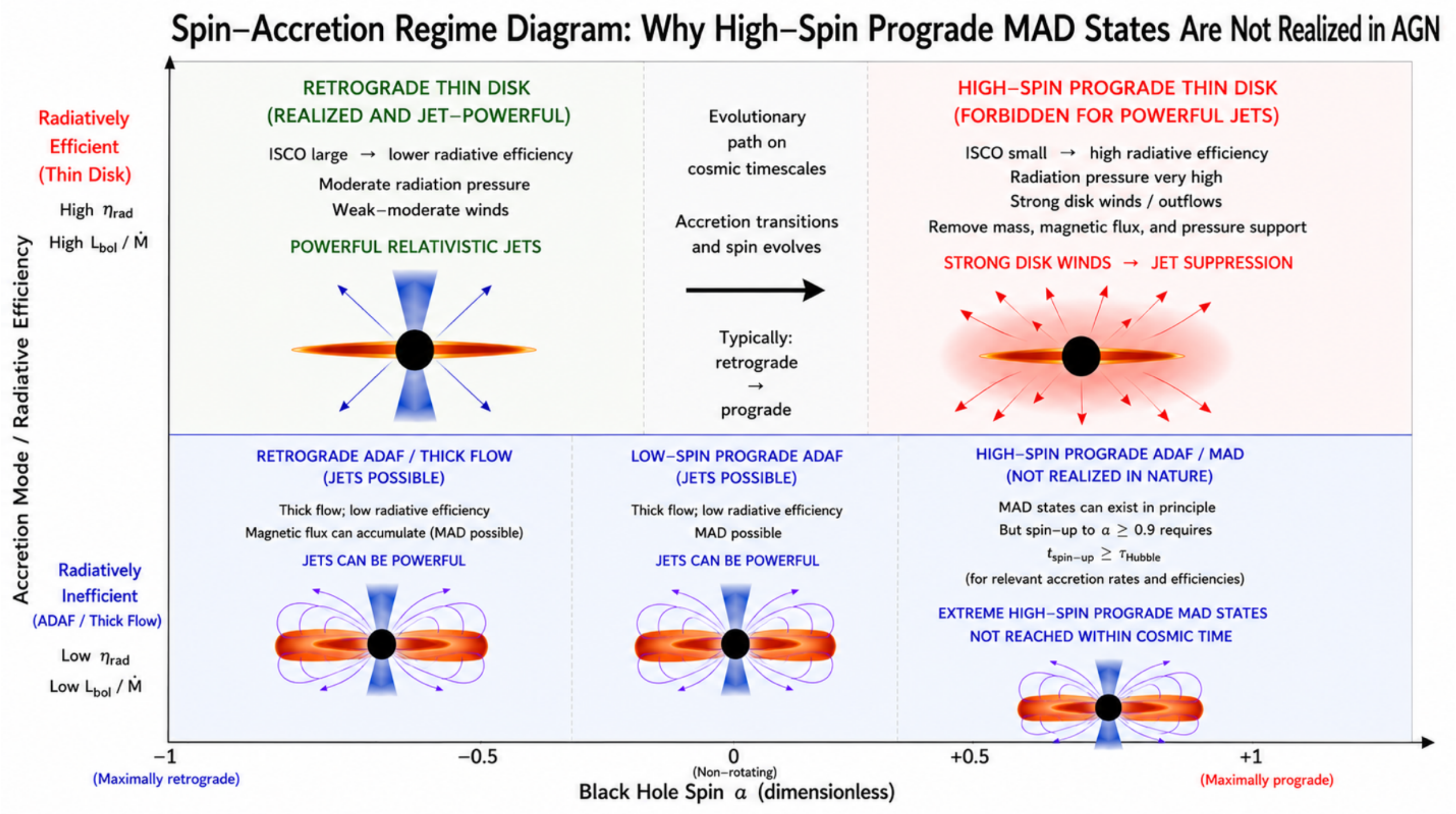

Figure 1: Schematic phase diagram of black hole accretion states as a function of spin and accretion mode. Powerful jets are produced in retrograde thin disks (top left) and in radiatively inefficient flows (bottom) over a range of spins: at moderate retrograde spins and at low-to-moderate prograde spins. High-spin prograde thin disks (top right) are inhibited by strong radiation driven winds, while high spin prograde MAD states (bottom right) are not reached within a Hubble time due to long spin-up timescales in radiatively inefficient accretion. This framework naturally explains the radio loud/radio quiet dichotomy and the cosmic evolution of powerful radio sources.

This interpretation removes the apparent tension between simulations and the counterrotation–corotation paradigm. Simulations identify what is physically possible under specific conditions, while the evolutionary framework determines which of those possibilities are realized. In this sense, the absence of high-spin prograde MAD states in nature is not a failure of either approach, but a natural consequence of their synthesis.

## 5. Physical Constraints on Accretion States

The preceding sections establish that numerical simulations identify a regime in which rapidly spinning black holes in prograde, magnetically arrested configurations produce highly efficient relativistic jets, while the counterrotation–corotation paradigm provides an evolutionary framework that successfully accounts for a wide range of observations. We now examine the physical constraints on accretion flows to determine which of these configurations are realized in astrophysical systems in AGN.

A central point is that accretion does not occur in a single, universal mode. Instead, black hole systems evolve through distinct accretion states characterized by different geometries, radiative properties, and transport processes. Two limiting regimes are relevant:

geometrically thin, radiatively efficient disks and geometrically thick, radiatively inefficient flows. As we show below, each of these regimes imposes constraints that prevent the realization of high-spin prograde magnetically arrested states.

### 5.1 Thin Disk Regime: Radiative Enhancement and Jet Suppression

For accretion flows associated with quasar-like and other bright AGN, the disk is expected to be geometrically thin. In this regime, the radiative efficiency is determined by the binding energy at the innermost stable circular orbit (ISCO),

$$\eta(a) = 1 - E_{\mathrm{ISCO}}(a),$$

with $a$ the dimensionless spin parameter, which increases strongly as the ISCO moves inward for increasing prograde spin. The local radiative flux in a thin disk can be expressed as

$$F(r) \sim \frac{\dot{M}}{r}\,\mathcal{F}(r,a),$$

where $F(r)$ is the radiative flux emitted per unit surface area at radius $r$, $\dot{M}$ is the mass accretion rate, and $\mathcal{F}(r,a)$ is a dimensionless function that encodes the relativistic corrections associated with the spacetime geometry and the black hole spin parameter $a$ with a strong enhancement toward the inner edge of the disk. To leading order, the peak flux scales approximately as

$$F_{\max} \propto \frac{\dot{M}}{r_{\mathrm{ISCO}}^2}.$$

Thus, as $r_{\mathrm{ISCO}}$ decreases with increasing prograde spin, the radiation field becomes increasingly intense and centrally concentrated. This increase in radiative output has direct dynamical consequences. The corresponding radiation pressure,

$$P_{\mathrm{rad}} \sim \frac{F}{c},$$

rises rapidly toward high prograde spin and naturally drives disk outflows. Observationally, such outflows are manifested as powerful winds, including ultra-fast outflows detected in X-ray spectra. These winds represent a significant channel for the removal of mass, energy, and angular momentum from the system.

The presence of strong disk-driven outflows fundamentally alters the conditions required for jet formation. Radiatively driven winds (i) disrupt the large-scale organization of magnetic fields, (ii) remove mass and magnetic flux from the inner disk, and (iii) compete with jet formation as an outlet for accretion energy. As a result, the conditions necessary to produce powerful, collimated relativistic jets are not maintained in high-spin prograde thin disks. This provides a natural explanation for the jet–wind anti-correlation discussed in Section 3. As radiative efficiency increases, wind power increases and jet production is suppressed. In this regime, high-spin prograde systems are therefore expected to be radio quiet.

### 5.2 Magnetically Arrested States and Accretion Mode

Magnetically arrested disk states require the accumulation of large-scale poloidal magnetic flux on the black hole to the point where magnetic pressure significantly influences the accretion flow. This process depends critically on the ability of the disk to advect magnetic flux inward without significant losses.

Such conditions are naturally satisfied in geometrically thick, radiatively inefficient flows, in which radial inflow speeds are relatively high and radiative losses are negligible. In contrast, thin disks are characterized by slower inflow, efficient radiation, and strong outflows, all of which act to limit the accumulation of magnetic flux.

In the context of the counterrotation–corotation paradigm, radiatively inefficient flows correspond to a later stage of accretion, often associated with advection-dominated accretion flows (ADAFs). These flows provide a natural environment for the formation of magnetically arrested states. However, they do not represent the conditions under which the highest radiative efficiencies—and thus the strongest thin-disk effects—are realized.

### 5.3 Spin Evolution and Timescale Constraints

The formation of high-spin prograde black holes requires sustained accretion over extended periods. The evolution of the dimensionless spin parameter $a$ under accretion can be expressed schematically as

$$\frac{da}{dt} \sim \dot{M},$$

with the precise rate depending on the angular momentum of the accreted material and relativistic corrections near the ISCO.

In radiatively efficient thin disks, spin-up can in principle proceed relatively rapidly. However, as shown above, this regime is associated with strong radiative output and disk-driven outflows that suppress jet formation. Thus, even if high prograde spin is achieved, the conditions required for powerful jet production are not present.

In contrast, radiatively inefficient flows provide conditions favorable for jet production and for the formation of magnetically arrested states. However, in this regime, accretion proceeds less efficiently in terms of angular momentum transport and radiative output, and the timescales required to spin the black hole up to near-maximal prograde rotation are correspondingly longer. For the relevant systems, these timescales are comparable to or exceed the age of the universe.

As a result, the two requirements for maximizing jet power in simulations—namely, high prograde spin and magnetically arrested accretion—are not simultaneously satisfied in realistic astrophysical systems.

The analysis above reveals a fundamental incompatibility between the physical conditions required to achieve high-spin prograde states and those required to sustain magnetically arrested accretion.

- In the thin disk regime, high prograde spin is associated with strong radiation and powerful disk winds, which suppress jet formation and prevent the conditions necessary for magnetically arrested states.
- In the radiatively inefficient regime, magnetically arrested states can form and support powerful jets, but the timescales required to reach near-maximal prograde spin are prohibitively long.

The configuration that maximizes jet power in simulations—high-spin prograde accretion in a magnetically arrested state—therefore requires the simultaneous realization of conditions that are mutually incompatible in astrophysical systems. It follows that such configurations are not realized in nature.

Instead, these states should be understood as defining the upper bounds of jet production under idealized conditions, rather than as representative of physically accessible systems. When combined with the evolutionary framework of the counterrotation–corotation paradigm, this conclusion provides a consistent explanation for the observed properties of active galactic nuclei.

5.4 Horizon-Scale Constraints from M87

M87 provides the most direct observational test of the accretion states discussed here because its horizon-scale emission has been resolved by the Event Horizon Telescope (EHT) and compared extensively with GRMHD simulations. The first EHT analysis of M87 compared the observed 230 GHz structure with a large library of GRMHD models spanning different black-hole spins, magnetic-flux states, and plasma prescriptions. The observations were consistent with emission produced in the immediate vicinity of a Kerr black hole, while non-spinning models failed (Event Horizon Telescope Collaboration 2019). Importantly, however, the EHT images did not provide a unique measurement of the black-hole spin. Rather, several combinations of spin, magnetic flux, and plasma properties were capable of reproducing important features of the horizon-scale emission.

The polarization observations provide a stronger constraint on the magnetic state of the accretion flow. Comparison of the polarized EHT images with GRMHD calculations substantially favor magnetically arrested disk models over weakly magnetized SANE models. The viable MAD models produce jet powers of approximately $10^{42}$–$10^{43}$ erg $s^{-1}$ and imply an accretion rate of approximately (3–20) × $10^{-4}$ $M_{\odot}$ $yr^{-1}$, corresponding to a radiatively inefficient accretion flow (Event Horizon Telescope Collaboration 2021). Thus, M87 provides observational support for the presence of dynamically important magnetic flux near the horizon and for a MAD-like state in a low-Eddington-ratio system. It does not, however, establish that this MAD state is associated with near-maximal prograde spin.

Independent attempts to constrain the spin illustrate the remaining model dependence. Broadband spectral and jet-morphology modeling with GRMHD simulations finds that approximately $0.5 \lesssim a \lesssim 1$ can reproduce important properties of the M87 jet, with higher prograde spins potentially favored in particular models (Fromm et al. 2022). Such results demonstrate that high-spin prograde MAD solutions are physically viable GRMHD configurations, but they do not constitute a unique measurement of the spin of M87. The inferred spin depends on assumptions concerning the magnetic state, electron distribution, radiative prescription, jet energetics, and relationship between the simulated flow and the observed emission.

More recent GRMHD calculations illustrate this degeneracy more clearly. Radiative two-temperature MAD simulations find a strong dependence of the polarization pitch angle on black-hole spin. When this observable is compared with the EHT measurements, the preferred range is approximately $-0.7 \lesssim a \lesssim 0.2$, although the authors explicitly caution that none of their models simultaneously satisfies all EHT polarization constraints (Chael 2025). The result therefore should not be interpreted as a measurement of retrograde spin. It demonstrates instead that horizon-scale observables can be reproduced within substantially different regions of spin parameter space depending upon the plasma physics included in the GRMHD calculation.

The most recent multi-epoch EHT analysis provides an additional constraint that is particularly relevant here. Comparison of the observed ring asymmetry with strongly magnetized GRMHD models marginally disfavors very small spins, $|a| \lesssim 0.2$, but does not determine the sign of the spin or require a near-maximal prograde configuration (Event Horizon Telescope Collaboration 2026). The emerging horizon-scale picture therefore favors a rotating black hole supplied with dynamically important magnetic flux while leaving a broad range of spin magnitude and orientation viable.

M87 consequently provides an important distinction between the existence of MAD accretion and the existence of an extreme high-spin prograde MAD state. The former is increasingly supported by horizon-scale observations; the latter has not been established observationally. This distinction is consistent with the evolutionary argument developed in this work. M87 is presently accreting at a very low Eddington ratio and therefore occupies precisely the radiatively inefficient regime in which MAD states and powerful jets are expected to be possible. The relevant question is therefore not whether M87 can be magnetically arrested, but whether its evolutionary history can produce a near-maximally spinning prograde black hole while it remains in this low-accretion state.

For a black hole of mass $M \simeq 6.5 \times 10^9 M_\odot$, as measured by the EHT, the small present-day accretion rate inferred from the polarized GRMHD models makes substantial spin evolution extremely slow. The current accretion state therefore cannot plausibly spin M87 from modest spin to near-maximal prograde rotation on a cosmologically relevant timescale. Any very large prograde spin inferred for M87 would consequently have to be inherited from an earlier high-accretion phase rather than generated during its present MAD phase. This provides a direct observational formulation of the evolutionary constraint proposed here. MAD accretion can readily occur in radiatively inefficient systems, but the existence of the MAD state itself does not demonstrate that the extreme high-spin prograde configurations used to maximize jet efficiency in GRMHD simulations are populated in nature.

The present horizon-scale observations of M87 therefore neither require nor provide direct evidence for the extreme high-spin prograde MAD state. Instead, they establish three more limited conclusions: M87 contains a rotating black hole, its near-horizon accretion flow is strongly magnetized and is well described by MAD models, and its spin remains model dependent. M87 is consequently consistent with the central distinction emphasized here between configurations that can be constructed in GRMHD simulations and those that are demonstrably realized through astrophysical evolution.

GRMHD does not uniquely identify FRII jets with retrograde accretion, but it demonstrates that high-spin retrograde configurations are fully capable of producing the required FRII jet powers. In particular, Tchekhovskoy & McKinney (2012) find a total outflow efficiency of 38 ± 4% for a retrograde black hole with spin $a = -0.9$, with approximately 80% of the extracted energy carried by the Blandford–Znajek-powered relativistic jet, corresponding to a jet efficiency of approximately 30%. For accretion rates of 0.1–1 solar masses per year, this implies jet powers of approximately $2 \times 10^{45}$– $2 \times 10^{46}$ erg $s^{-1}$, squarely within the range of powerful FRII radio sources. Thus, assigning powerful FRIIs to retrograde black holes is quantitatively consistent with GRMHD; the additional assumption of the present framework is that the still more efficient high-spin prograde MAD solutions permitted by GRMHD are not realized in astrophysical systems for AGN (Tchekhovskoy & McKinney 2012, MNRAS, 423, L55).

5.5 X-ray binaries

X-ray binaries provide an instructive comparison because they can access magnetically arrested accretion states without requiring the counterrotation–corotation evolution invoked here for AGN. In our framework, X-ray binary disks remain corotating, so their rapid transitions between jet-active and jet-suppressed states are driven primarily by changes in accretion rate and disk structure. At low accretion rates, a geometrically thick, radiatively inefficient inner flow can accumulate substantial magnetic flux and sustain a MAD-like, jet-producing state. As the accretion rate rises toward the Eddington regime due to changes provided by the donor star, the flow becomes geometrically thin and radiatively efficient and extends inward toward the ISCO. For high prograde spin, the increased thin-disk efficiency produces a strong radiation field and disk outflows that suppress the jet. Thus, X-ray binaries can access MAD states, but their much smaller black-hole masses and correspondingly smaller absolute accretion rates prevent them from reaching the enormous absolute jet powers associated with the most powerful MAD configurations representing powerful radio loud AGN in GRMHD simulations.

## 6. Discussion and Conclusions

In this work, we have examined the relationship between numerical simulations of jet production and the observed properties of active galactic nuclei within the framework of the counterrotation–corotation paradigm. While simulations demonstrate that rapidly spinning black holes in prograde, magnetically arrested states can produce highly efficient relativistic jets, we have argued that the physical conditions required for such configurations are not realized in astrophysical systems.

The key result of this paper is that the two requirements for maximizing jet power in simulations—namely, high prograde spin and magnetically arrested accretion—arise in distinct and mutually incompatible accretion regimes. In geometrically thin disks associated with quasar-like AGN, increasing prograde spin leads to enhanced radiative efficiency and strong disk-driven outflows, which suppress jet formation and prevent the accumulation of magnetic flux required for magnetically arrested states. Conversely, in geometrically thick, radiatively inefficient flows capable of sustaining magnetically arrested accretion, the timescales required to spin black holes up to near-maximal prograde rotation are comparable to or exceed the age of the universe. As a result, the extreme high-spin prograde configurations that maximize jet power in simulations are not physically accessible.

This conclusion removes the apparent tension between simulation results and the counterrotation–corotation paradigm. The latter does not require that retrograde accretion produce the most powerful jets, but only that it provide a regime in which powerful jets are efficiently generated and sustained. This is sufficient to support the time-dependent spin evolution that links different classes of active galactic nuclei and explains the observed decline of powerful radio sources from high redshift to the present epoch.

Within this unified framework, the radio-loud/radio-quiet dichotomy emerges naturally from the interplay between spin orientation and accretion state. Retrograde systems, particularly at earlier stages of evolution, are associated with powerful jet production, while prograde thin-disk systems correspond to the radio-quiet population characterized by strong radiative output and disk-driven winds. The observed anti-correlation between jets and winds follows directly from the competing roles of these processes as channels for the release of accretion energy.

More broadly, this work highlights the importance of distinguishing between physically possible configurations identified in simulations and those that are realized through the evolutionary pathways of astrophysical systems. Numerical simulations define the limits of jet production under specific conditions, while the combined constraints of accretion triggers and black hole evolution determine which of these limits are accessible in nature.

We therefore conclude that high-spin prograde magnetically arrested configurations, while valid solutions of the governing equations, do not describe astrophysical systems for AGN. When this constraint is considered, numerical simulations and the counterrotation–corotation paradigm are fully compatible and together provide a coherent and comprehensive framework for understanding jet production and the radio-loud/radio-quiet dichotomy in active galactic nuclei.

Acknowledgments

I thank the referee for noticing an issue that required extending the discussion to X-ray binaries.